\documentclass[fleqn,usenatbib]{mnras}
\usepackage{newtxtext,newtxmath}
\usepackage[T1]{fontenc}
\DeclareRobustCommand{\VAN}[3]{#2}
\let\VANthebibliography\thebibliography
\def\thebibliography{\DeclareRobustCommand{\VAN}[3]{##3}\VANthebibliography}

\usepackage[dvipdfmx]{graphicx}	
\usepackage{amsmath}	
\usepackage{color}
\usepackage{threeparttable}

\newcommand{\msun}{{\rm M_\odot}} 
\newcommand{\yr}{{\rm yr}} 
\newcommand{\pc}{{\rm pc}} 

\newcommand{\femg}{{\rm [Fe/Mg]}}
\newcommand{\fealpha}{{\rm [Fe/\alpha]}}
\newcommand{\ms}{M_\ast}
\newcommand{\mbh}{M_\bullet}
\newcommand{\mmin}{{M_{\rm min}}}
\newcommand{\mmax}{{M_{\rm max}}}
\newcommand{\yfe}{Y_{\rm Fe}}
\newcommand{\ymg}{Y_{\rm Mg}}

\newcommand{\mdot}{\dot{M}_\ast}
\newcommand{\mdotmin}{\dot{M}_{\rm 0.1}}
\newcommand{\mc}{M_{\rm c}}

\defcitealias{Nomoto2013ARA&A}{NKT13}
\defcitealias{Heger2010ApJ}{HW10}
\defcitealias{Umeda2008ApJ}{UN08}

\title[Top-heavy IMF in galactic nuclei]{Top-heavy stellar mass distribution in galactic nuclei inferred from the universally high abundance ratio of [Fe/Mg]}

\author[D.~Toyouchi et al.]{
Daisuke~Toyouchi$^{1}$\thanks{E-mail: toyouchi@resceu.s.u-tokyo.ac.jp},
Kohei~Inayoshi$^{2}$,
Miho~N.~Ishigaki$^{3}$,
and Nozomu~Tominaga$^{3,4,5}$
\\
$^{1}$Research Center for the Early Universe (RESCEU), The University of Tokyo,
Hongo, 7-3-1, Bunkyo-ku
Tokyo, 113-0033, Japan\\
$^{2}$Kavli Institute for Astronomy and Astrophysics, Peking University, Beijing 100871, China\\
$^{3}$National Astronomical Observatory of Japan, 2-21-1 Osawa, Mitaka, Tokyo 181-8588, Japan\\
$^{4}$Department of Physics, Faculty of Science and Engineering, Konan University, 8-9-1 Okamoto, Kobe, Hyogo 658-8501, Japan\\
$^{5}$Kavli Institute for the Physics and Mathematics of the Universe (WPI), The University of Tokyo, 5-1-5 Kashiwanoha, Kashiwa, Chiba 277-8583, Japan}

\date{Accepted XXX. Received YYY; in original form ZZZ}

\pubyear{2021}

\begin{document}
\label{firstpage}
\pagerange{\pageref{firstpage}--\pageref{lastpage}}
\maketitle

\begin{abstract}

Recent observations of active galactic nuclei (AGNs) have shown a high Fe~II/Mg~II line-flux ratio in their broad-line regions,
nearly independent of redshift up to $z \gtrsim 6$.
The high flux ratio requires rapid production of iron in galactic nuclei to reach an abundance ratio of $\femg \gtrsim 0.2$ as high as those observed in matured galaxies in the local universe.
We propose a possible explanation of rapid iron enrichment in AGNs by massive star formation that follows 
a top-heavy initial mass function (IMF) 
with a power-law index of $\Gamma$ 
larger than the canonical value of $\Gamma=-2.35$ for a Salpeter IMF. 
Taking into account metal production channels from different types of SNe, we find that the high value of $\femg \gtrsim 0.2$ 
requires the IMF to be characterized with $\Gamma \gtrsim -1$ ($\Gamma \gtrsim 0$) and a high-mass cutoff 
at $\mmax \simeq 100$--$150~\msun$ $(\mmax \gtrsim 250~\msun)$.
Given the conditions, core-collapse SNe with $\ms \gtrsim 70~\msun$ and pair-instability SNe give a major contribution for iron enrichment.
Such top-heavy stellar IMFs would be a natural consequence from mass growth of stars formed in dense AGN disks under Bondi-like gas accretion 
that is regulated by feedback at $\ms \gtrsim 10~\msun$.
The massive stellar population formed in AGN disks also leave stellar-mass black hole remnants, whose mergers associated with 
gravitational-wave emission account for at most 10 \% of the merger rate inferred from LIGO/Virgo observations to simultaneously explain the high $\femg$ ratio with metal ejection.
\end{abstract}




\section{Introduction}\label{sec:intro}

The existence of quasars and active galactic nuclei (AGNs) provides solid evidence of mass accretion onto 
supermassive black holes (SMBHs) \citep[][]{Lynden-Bell1969Natur, Ho2008ARA&A}.
Accretion disks powering bright AGNs are expected to be unstable by their self-gravity in the outer region at $\gtrsim 0.01$~pc 
and a good fraction of the gas is consumed by star formation 
\citep[][]{Paczynski1978AcA, Shlosman1987Natur, Goodman2003MNRAS, Inayoshi2016ApJ}.
Theoretical studies have suggested that stars embedded in the self-gravitating regimes of AGN disks undergo gas accretion 
after their formation and thus the stellar mass distribution tends to be top-heavy 
\citep[][]{Goodman2004ApJ, Nayakshin2007MNRAS, Mapelli2012ApJ}.
In fact, observations of the central nuclear region ($\sim 0.5$~pc) of the Milky Way have shown the existence of hundreds of 
massive, young OB stars and Wolf-Rayet stars \citep[][]{Krabbe1995ApJ, Genzel2003ApJ, Ghez2003ApJ, Levin2003ApJ},
and the absence of low-mass stars compared to the number expected with a standard Salpeter initial mass function (IMF) \citep[][]{Nayakshin2005MNRAS, Paumard2006ApJ, Alexander2007ApJ, Bartko2010ApJ}.

One possible way to study the stellar mass function in nuclear regions of external galaxies is to measure chemical abundance 
of interstellar media (ISM) in AGNs since the production efficiency of heavy elements depends on the progenitor mass sensitively.
Observations of high-$z$ quasars have extensively measured the line flux ratio of FeII and MgII emitted from the broad-line regions (BLRs) as a proxy of the abundance ratio of iron to $\alpha$ elements, $\fealpha$ \citep[e.g.,][]{Dietrich2002ApJ, Iwamuro2002ApJ, Barth2003ApJ, Freudling2003ApJ, Kurk2007ApJ}.
In general, $\fealpha$ is expected to decline toward higher redshifts because $\alpha$ elements are predominantly produced in core-collapse supernovae (CCSNe), whereas iron is mainly released in type Ia SNe with a long delay time of $\gtrsim$ 0.5~Gyr \citep[e.g.,][]{Matteucci1986A&A, Matteucci2001ApJ, Maoz2010ApJ, Homma2015ApJ}.
However, quasar observations show that the Fe~II/Mg~II flux ratio does not decrease but appears to be almost constant up to $z \sim 7.5$, 
by which type Ia SNe hardly make a contribution to iron production
\citep{Iwamuro2002ApJ, Dietrich2003ApJ, Maiolino2003ApJ, De_Rosa2011ApJ, Mazzucchelli2017ApJ, Shin2019ApJ, Onoue2020ApJ, Schindler2020ApJ, Yang2021arXiv}.
Moreover, \citet{Sameshima2017ApJ, Sameshima2020ApJ} have extensively analyzed the observed spectra of high-z quasars and found that the high Fe~II/Mg~II flux ratio is caused by overproduction of iron, namely an extremely high iron to magnesium abundance ratio 
$\femg > 0.2$.
Therefore, those observations require the early production of iron in AGNs through a pathway different from  normal CCSNe and type Ia SNe.

We propose a possible explanation to understand the origin of fully-established chemical environments in AGNs at a wide range of redshift, 
considering a top-heavy mass distribution of stars born in AGNs.
This hypothesis leads to the frequent occurrence of massive CCSNe and pair-instability SNe (PISNe) and thus increases the $\femg$ abundance ratio in the ISM because energetic explosions caused by heavier progenitors generally provide a larger amount of iron relative to magnesium in their ejecta.
Such short-lived stellar progenitors establish iron-enriched BLRs in a short time of $<$ 10~Myr, which would naturally explain the apparent universality of the high flux ratio of Fe~II/Mg~II.
In this paper, we quantify the shape of the stellar mass distribution that reproduces $\femg > 0.2$ as observed in quasar BLRs.
Moreover, if a substantial fraction of those massive stars end up in binary BHs in galactic nuclei, their mergers would be detectable as gravitational-wave sources for both ground- and space-based interferometers such as advanced LIGO/Virgo and LISA/Tianqin.

The rest of the paper is organized as follows.
We first introduce the stellar yields we assume in Section~\ref{sec:yield}, and present the shape of the stellar IMF that explains 
the observed abundance ratio of $\femg$ in Section~\ref{sec:femg}.
In Section~\ref{sec:how}, we model the history of mass growth of stars embedded in AGN disks and argue how the required 
top-heavy IMFs are established.
In Section~\ref{sec:bhmr}, the cosmic rate of binary BH mergers is calculated based on the top-heavy IMF hypothesis.
In Section~\ref{sec:other}, we briefly discuss alternative explanations of the high-$\femg$ ratios observed in BLRs.
Finally, our conclusion is summarized in Section~\ref{sec:summary}.
Throughout this paper, we adopt $\femg = 0.2$ as a reference value of the observed abundance ratio in quasars.


\section{Implication of top-heavy IMFs}\label{sec:thimf}

\subsection{Stellar Yields}\label{sec:yield}

\begin{figure}
\begin{center}
\includegraphics[width=\columnwidth]{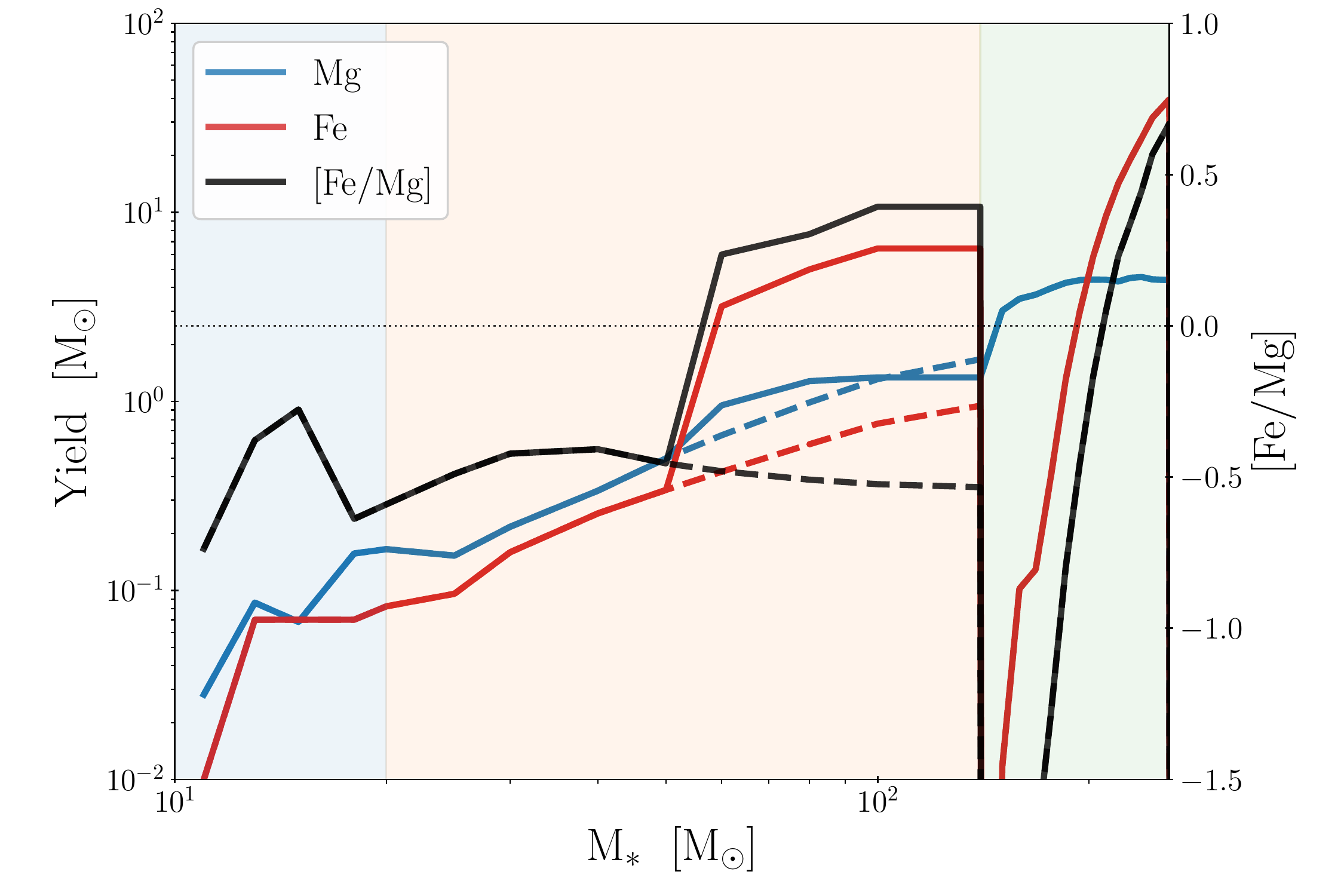}
\end{center}
\caption{
The amount of stellar yields as a function of stellar progenitor mass 
(see also Table \ref{table:yield}).
Each curve represents the yield mass of iron (red) and magnesium (blue) and the abundance ratio of $\femg$ in ejecta (black).
The solid and dashed curves at $\ms = 50$--$140~\msun$ indicate 
the stellar yields with and without a contribution from SLSN-like events, respectively (see Section~\ref{sec:yield} for more details).
The shaded region shows the mass range where CCSNe, HNe/SLSNe, and PISNe dominate the stellar yield.}
\label{fig:yield}
\end{figure}

The amount of heavy elements produced by a SN event considerably 
depends on the mass of its stellar progenitor.
In this paper, to examine the chemical enrichment process led by stellar populations with a top-heavy IMF, 
we consider SN yields in a wide range of mass over $10\leq \ms /\msun \leq 260$.
Figure~\ref{fig:yield} shows the yield mass of iron and magnesium as a function of progenitor stellar mass 
taken from previous studies in the literature.
We adopt the yield tables calculated for zero-metal stars, which hardly suffer from mass loss by stellar winds.
Additionally, as mentioned in Section~\ref{sec:intro}, we do not consider metal production from type Ia SNe 
because of their delay timescales since their progenitors form ($\simeq 1$ Gyr), which is comparable or even 
longer than the cosmic age at $z\gtrsim 6$ when high-redshift quasars have been iron-enriched.
Metal loading processes through stellar winds from giant stars and mergers of neutron stars are not considered
because of their minor contribution to iron and magnesium abundances.

Massive stars with mass of $10 \leq \ms/\msun \leq 140$ end up as CCSNe at the end of their stellar lifetime.
Stars with $\ms = 10-20~\msun$ cause normal CCSNe with explosive energies of $E_{\rm SN} \simeq 10^{51}~{\rm erg}~(\equiv E_{51})$ 
and leave neutron stars behind (the blue region in Figure~\ref{fig:yield}).
We assume those normal CCSNe to produce yields calculated by 
\citepalias[Nomoto, Kobayashi \& Tominaga 2013, hereafter][]{Nomoto2013ARA&A}.
Stars with $\ms = 20-140~\msun$ are considered to leave BHs behind their explosions (the orange region in Figure~\ref{fig:yield})
and are often observationally classified into two groups: 
faint SNe and hyper novae (HNe).
Faint SNe would be failed explosion events with $E_{\rm SN} < 10^{51}~{\rm erg}$, in which a large fraction of the central iron core falls 
back into the remnant BH.
In contrast, HNe are very energetic events with $E_{\rm SN} > 10^{52}~{\rm erg}$ and tend to be associated with long 
gamma-ray bursts (GRBs) that are caused by relativistic jets produced by rapidly spinning BHs in massive stellar collapsars.
The relative contribution of the two types of explosion to metal production is highly uncertain and poorly constrained both from observations and theoretical studies. 
Nevertheless, we adopt the yield model based on a hypothesis that all stars in the mass range explode as HNe because 
faint SNe provide a tiny mass of iron and cannot reproduce the high abundance ratio of $\femg$ observed in BLRs.
This assumption would be reasonable because stars formed in AGN disks are expected to be spun up due to 
strong shear motion in the disks \citep{Jermyn2021ApJ}.

\begin{figure}
\begin{center}
\includegraphics[width=\columnwidth]{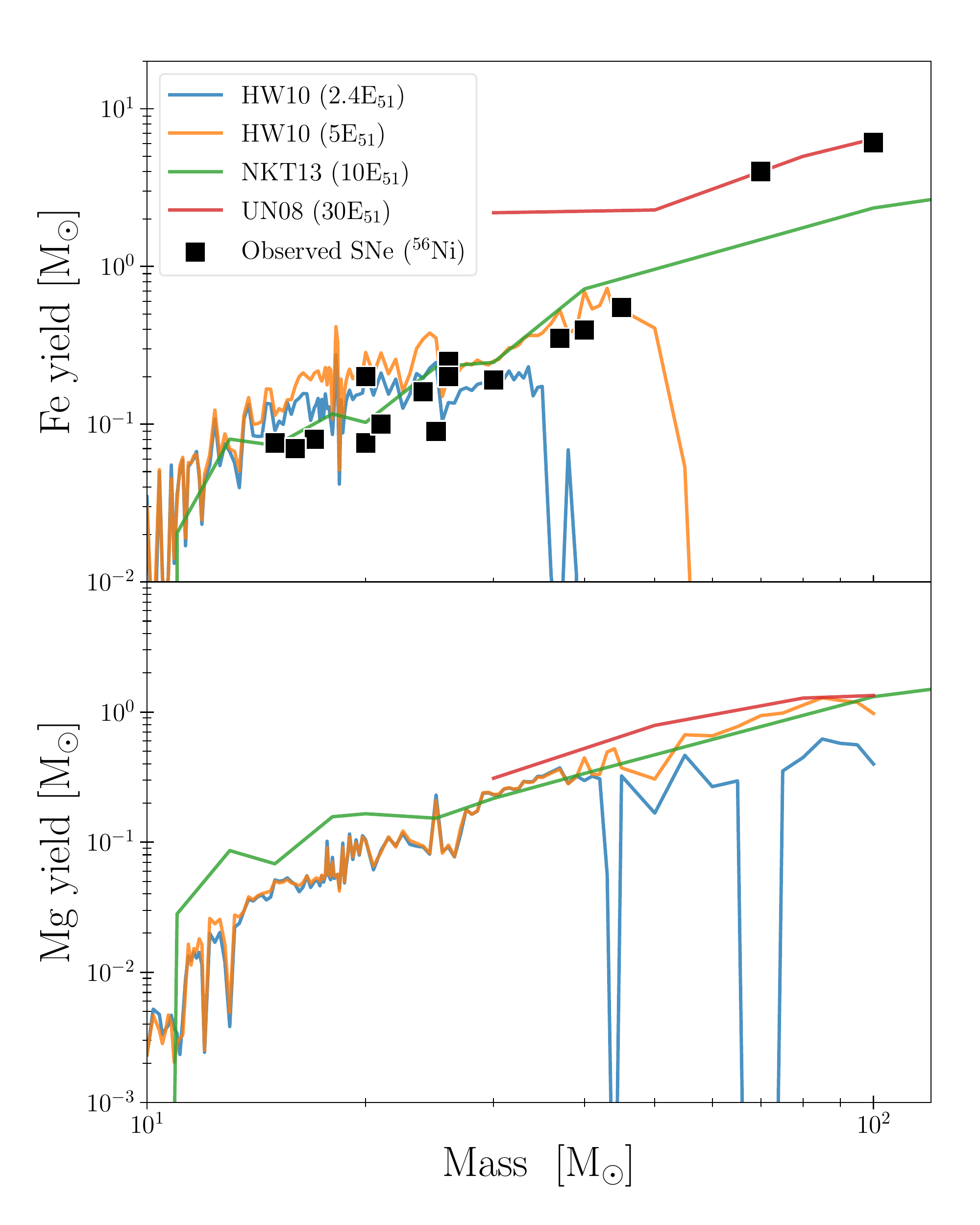}
\end{center}
\caption{
Summary of the stellar yield models for iron (top) and magnesium (bottom) for different explosive energies of $E_{\rm SN}/E_{51} = 2.4$ and $5.0$  \citep{Heger2010ApJ}, $10$ \citep{Nomoto2013ARA&A}, and $30$ \citep{Umeda2008ApJ}.
For comparison, we overlay the $\rm ^{56}Ni$ mass measured from observed SN events in the top panel (see the text for more details).
Mass ejection from the most energetic explosion (red curve), where no fallback of synthesized metals to the remnant compact object is assumed, is required to explain the $\rm ^{56}Ni$ mass synthesized in the observed SLSNe.
}
\label{fig:yield_HN}
\end{figure}

\begin{figure*}
\begin{center}
\includegraphics[width=2\columnwidth]{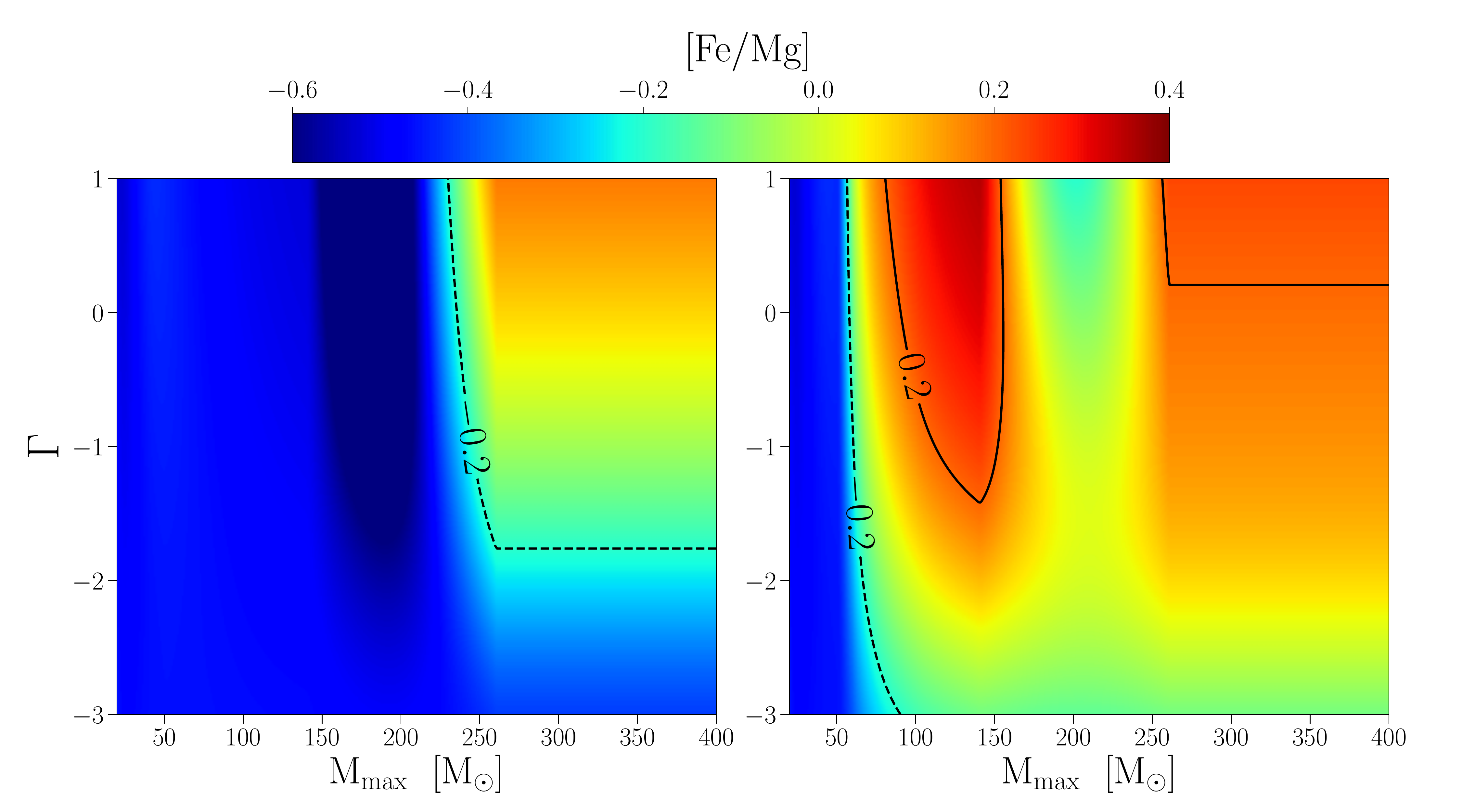}
\end{center}
\caption{
The abundance ratio of iron to magnesium evaluated with Eq.~(\ref{eq:femg}) for two parameters that characterize stellar IMF; the maximum mass $\mmax$ and the IMF slope $\Gamma$.
The left and right panels show the value of $\femg$ based on the yield model without and with a contribution from SLSN-like events.
The solid-contours indicate $\femg = 0.2$, the abundance ratio observed in BLRs of quasars, which we attempt to reproduce in this paper.
For comparison, the dashed-contour denotes $\femg = -0.2$, which corresponds to the upper bound found for the Galactic metal-poor stars.
}
\label{fig:y_fe_mg}
\end{figure*}

The predicted amount of stellar yields produced by HNe events depends on the explosion energies and mass-fall efficiency.
Figure~\ref{fig:yield_HN} shows the iron and magnesium yield mass obtained from the following four models with different explosion energies:
$E_{\rm SN} = 2.4~E_{51}$ and $5.0~E_{51}$ \citepalias[Heger \& Woosley 2010, hereafter][]{Heger2010ApJ}, 
$E_{\rm SN} = 10~E_{51}$ \citepalias{Nomoto2013ARA&A}, 
and $E_{\rm SN} = 30~E_{51}$ \citepalias[Umeda \& Nomoto 2008, hereafter][]{Umeda2008ApJ}.
All the four models make a robust prediction of the amount of magnesium yields in SN ejecta.
This is because the magnesium-yield mass is determined by nuclear synthesis in the pre-SN phase rather than that in the SN ejecta.
On the other hand, the iron yield mass highly depends on the explosion energy at $\ms \gtrsim 50~\msun$ and 
increases with the explosion energy. 
For the most energetic case with $E_{\rm SN}=30~E_{51}$, the iron yield mass becomes substantially 
higher than the other models because all the iron synthesized in the exploding core is loaded in the ejecta, 
i.e., no fallback of the iron onto the remnant BH.

In the top panel of Figure 2, we overlay the iron yield mass estimated from the observed light curves of SNe by assuming that
the iron mass equals to the ${\rm ^{56}Ni}$ mass released in the ejecta 
(the data taken from \citealt{Hashimoto1989A&A, Iwamoto2000ApJ, Hatano2001AAS, Maeda2002ApJ, Mazzali2002ApJ, Mazzali2003ApJ, Mazzali2010MNRAS, Moriya2010ApJ}, and Figure~15 of \citealt{Mazzali2017MNRAS}).
For SN events produced by progenitors with $\ms \lesssim 50~\msun$, the observed ${\rm ^{56}Ni}$ mass is well explained 
by the yield models with low explosion energies of $E_{\rm SN} \leq 10~E_{51}$.
However, two energetic events of SN 1999as and SN 2007bi (the two rightmost data in Figure~\ref{fig:yield_HN}) are expected to
produce high ${\rm ^{56}Ni}$ mass of $\sim 5~\msun$.
This high ${\rm ^{56}Ni}$ mass can be explained by the yield model of \citetalias{Umeda2008ApJ}, where a high explosion energy 
and no fallback of synthesized metals are considered \citep[see also][]{Moriya2010ApJ, Moriya2019MNRAS}.
The two energetic SN events are the so-called super-luminous supernovae (SLSNe) since their peak luminosities are 
$\gtrsim 10$ times brighter than the typical value of normal CCSNe or HNe\footnote{SN 1999as and SN 2007bi are 
representative SLSN events whose light curves are well explained by extremely massive collapsar models.
However, we note that observed light curves of SLSNe show a diversity, which might be caused by various energy sources of those explosions, 
such as rapid spin-down of remnant magnetars and/or interaction between SN ejecta with dense circum-stellar media 
\citep[see][, references thereein]{Gal-Yam2019ARA&A}.
}.
%
Motivated by those observational facts, we here adopt the iron and magnesium yield models summarized in Figure~\ref{fig:yield} (solid curves);
\citetalias{Nomoto2013ARA&A} at $\ms = 20-50~\msun$ and \citetalias{Umeda2008ApJ} at $\ms = 60-140~\msun$, 
supposing HN and SLSN-like events, respectively, and stellar yields at $\ms = 50-60~\msun$ are calculated by a linearly interpolation.
In this case, we expect high iron to magnesium abundance in the SN ejecta, $\femg \sim 0.3$, around $\ms \sim 100~\msun$.
For comparison with this model that takes into account the SLSN-like yield, we also consider the case assuming the HN yield of \citetalias{Nomoto2013ARA&A} in the whole mass range of $\ms = 20-140~\msun$, which is also shown with the dashed line in Figure~\ref{fig:yield}.

Finally, in the highest mass range of $140 < \ms/\msun \leq 260$ (the green region in Figure~\ref{fig:yield}), we adopt a yield model 
for PISNe \citep{Heger2002ApJ}.
Since PISNe disrupt the entire structure of their stellar progenitors, a larger amount of heavy elements are ejected into 
the surrounding media compared to those by CCSNe and HNe.
The synthesized mass of iron increases with the progenitor mass and reaches $40~\msun$ for a star with $\ms = 260~\msun$, 
which corresponds to $\femg \simeq 0.6$.
Note here that while the PISN yield model we adopt does not take into account stellar rotation and magnetic fields, those effects are expected not to make a significant difference in iron and magnesium production \citep[][]{Takahashi2018ApJ}.

Beyond $\ms = 260~\msun$, we assume all the synthesized metals in stellar cores fall back to the remnant BHs and 
thus neglect mass ejection of iron and magnesium.
However, those massive stars could eject metals if GRB-like explosions are triggered in their stellar collapsars \citep{Ohkubo2006ApJ}.
In Section~\ref{sec:other}, we briefly mention the impact of massive GRBs on enhancement of $\femg$.

\subsection{IMFs required to explain the high $\femg$ ratio}\label{sec:femg}

Throughout this paper, we assume a stellar IMF to be characterized with a single-power law function of $\phi \propto m^\Gamma$.
The normalization of the IMF is set as
\begin{eqnarray}
\int^{\mmax}_{\mmin} m \phi(m) {\rm d}m = 1~\msun \ ,
\label{eq:imf}
\end{eqnarray}
where $\mmin$ and $\mmax$ are the minimum and maximum mass of stars, respectively.
We adopt $\mmin = 0.1~\msun$ because the estimated value of $\femg$ hardly depends on $\mmin$ as long as $\mmin<10~\msun$ for $\Gamma$ and $\mmax$ of interest.
Combined with the yield model shown in Figure~\ref{fig:yield}, we calculate the mass-integrated yields of iron and magnesium as
\begin{eqnarray}
Y_i = \int^{\mmax}_{\mmin} y_i(m) \phi(m) {\rm d}m \ ,
\label{eq:yield}
\end{eqnarray}
where the subscript $i$ represents ${\rm Fe}$ and ${\rm Mg}$.
For comparison with observations, we introduced the abundance ratio $\femg$ defined by 
\begin{eqnarray}
\femg = {\rm log}_{10}\left(\frac{\yfe}{\ymg} \frac{A_{\rm Mg}}{A_{\rm Fe}}\right) - {\rm log}_{10}\left(\frac{N_{\rm Fe}}{N_{\rm Mg}}\right)_\odot \ ,
\label{eq:femg}
\end{eqnarray}
where the mass number of iron and magnesium isotopes is $A_{\rm Fe} = 56$ and $A_{\rm Mg} = 24$, respectively, and 
the second term in the right hand side is the solar abundance ratio, as which we refer to the data of \citet{Asplund2005ASPC} 
in this paper.

Figure~\ref{fig:y_fe_mg} shows the abundance ratio calculated for different values of $\mmax$ and $\Gamma$.
We here overlay two contours of $\femg = -0.2$ (dashed) and $\femg=0.2$ (solid), 
each of which corresponds to the upper bound for metal-poor stars in our Galaxy \citep[e.g.,][]{Frebel2015ARA&A, Yong2021MNRAS} and 
the typical abundance ratio in BLRs of quasars, respectively.
Without yields produced by SLSN-like explosions (the left panel of Figure~\ref{fig:y_fe_mg}), the abundance ratio of $\femg$ is 
lower than the value for Galactic metal-poor stars at $\mmax < 230~\msun$ or $\Gamma <-1.8$.
The abundance ratio increases due to efficient iron production by PISNe at $\mmax > 250~\msun$ but does not reach $\femg = 0.2$.
On the other hand, including SLSN-like yields (the right panel of Figure~\ref{fig:y_fe_mg}), the iron abundance is efficiently enhanced 
and reaches $\femg > 0.2$ seen in quasars at $\Gamma \gtrsim -1$ and $\mmax \sim 100-150~\msun$. 
In this case, another solution that reproduces $\femg > 0.2$ appears at $\Gamma \gtrsim 0$ and $\mmax \gtrsim 250~\msun$ owing to an extra enrichment by PISNe.
We also note that the ratio of $\femg$ for the latter solution depends only on $\Gamma$ as long as $\mmax > 260~\msun$ 
because those heavy stars directly collapse to BHs without metal production.

In the following discussion, we focus on the case with SLSN-like yields and consider two reference models of IMFs that 
produce $\femg \gtrsim 0.2$; namely, the model parameters are set to ($\mmax$, $\Gamma$) = ($130~\msun$, $-0.5$) and ($300~\msun$, $0.5$).
Each IMF model represents the low and high-$\mmax$ solution in Figure~\ref{fig:y_fe_mg} and is referred as the IMF-A and IMF-B, respectively.

\begin{table*}
\begin{center}
\caption{Our reference models of top-heavy IMFs that give $\femg > 0.2$.}
\label{table:model}
\begin{tabular}{ccccccc} \hline \hline
Model & $(\mmax,~\Gamma)$ & $\mdotmin$ & $\mc$ & $\alpha$ & $f_{\rm bh}$ & $f_{\rm pisn}$ \\
 & & $[{\rm 10^{-8}~\msun~\yr^{-1}}]$ & $[\msun]$ & & $[\msun^{-1}]$ & $[\msun^{-1}]$ \\
\hline
A & $(130~\msun,~-0.5)$ & $1.3 \pm 0.4$ 
& $14.3 \pm 4.3$ & $-1.4 \pm 0.5$ & $1.2 \times 10^{-2}$ & $0.0$ 
\\ \hline
B & $(300~\msun,~0.5)$ & $2.3 \pm 0.1$ & 
$9.4 \pm 5.3$ & $-0.5 \pm 0.5$ & $2.7 \times 10^{-3}$ & $2.7 \times 10^{-3}$ \\
 \hline \hline \\
\end{tabular}
\end{center}
\end{table*}


\section{Formation of top-heavy IMF in AGN disks}\label{sec:how}

Luminous AGNs are powered by mass accretion through dense gaseous disks, where newly-born stars grow in mass via continuous gas accretion.
In this section, we model the growth history of stars embedded in AGN disks so that their mass distribution explains the high $\femg$ abundance 
ratio observed in AGNs (see Section \ref{sec:femg}).

The time evolution of the stellar mass distribution is calculated 
by solving the continuity equation of 
\begin{eqnarray}
\frac{\partial \phi}{\partial t} + \frac{\partial}{\partial \ms} (\phi \mdot) = 0 \ ,
\label{eq:cont}
\end{eqnarray}
where the total number of stars is fixed (i.e., the source term in the right hand side is set to zero) and $\mdot$ is the growth rate of a star with a mass of $\ms$.
The growth rate of stars embedded in an AGN disk has not been understood properly owing to the uncertainties of stellar feedback (radiation and/or winds) and its impacts on the disk structure  
\citep[][]{Cantiello2021ApJ, Dittmann2021arXiv, Jermyn2021ApJ}.
To avoid numerous uncertainties in modeling of those effects,
we characterize the mass accretion rate using a simple functional form
with three free parameters:
\begin{eqnarray}
\mdot = \mdotmin \left( \frac{\ms}{0.1~\msun} \right)^2 \left( 1 + \frac{\ms}{\mc} \right)^{\alpha-2} \ ,
\label{eq:mdot}
\end{eqnarray}
where $\mdotmin$ is the mass growth rate at $\ms = 0.1~\msun$.
In this model, low-mass stars with $\ms \ll \mc$ undergo the Bondi-like accretion, 
$\mdot \propto \ms^2$, whereas for high mass stars with $\ms \gg \mc$, 
the mass growth rate is moderated to $\ms^{\alpha}~(\alpha < 2)$.
This modification is motivated by the effects suppressing stellar growth owing to stellar radiation/mechanical feedback 
($\alpha \simeq 1$; Eddington-limited growth), and the disk rarefication, gas shear motion, and tidal effects ($\alpha \lesssim 1$).

We initiate stars whose mass distribution follows a Salpeter IMF characterized with $\mmax = 50~\msun$ and $\Gamma = -2.35$.
The maximum mass of the IMF is adopted so that the ratio of $\femg$ becomes consistent with the upper bound for metal-poor stars in our Galaxy, $\femg \simeq -0.2$ (see also the right panel of Fig.~\ref{fig:y_fe_mg}).
We evolve the IMF by solving Eq.~(\ref{eq:cont}) for given three parameters and terminate its evolution at $t=5$ Myr, which 
corresponding to the lifetime of massive stars.
We evaluate the difference between the resultant mass distribution ($\phi_{\rm cal}$) and a reference top-heavy IMF ($\phi_{\rm TH}$)
that produces $\femg \simeq 0.2$, calculating the $\chi^2$ value defined by
\begin{eqnarray}
\chi^2 = \sum_i \left \{ {\rm log} \left ( \frac{\phi_{{\rm cal}, i}}{\phi_{{\rm TH},i}} \right ) \right \}^2 \ ,
\label{eq:chi2}
\end{eqnarray}
where the subscript $i$ indicates the $i$-th mass bin ($0.5\leq M_i/\msun \leq 500$).
We find the best-fitted parameter set of ($\mdotmin$, $\mc$ and $\alpha$) to minimize the $\chi^2$ value with the 
Markov Chain Monte Carlo (MCMC) method.

\begin{figure}
\begin{center}
\includegraphics[width=\columnwidth]{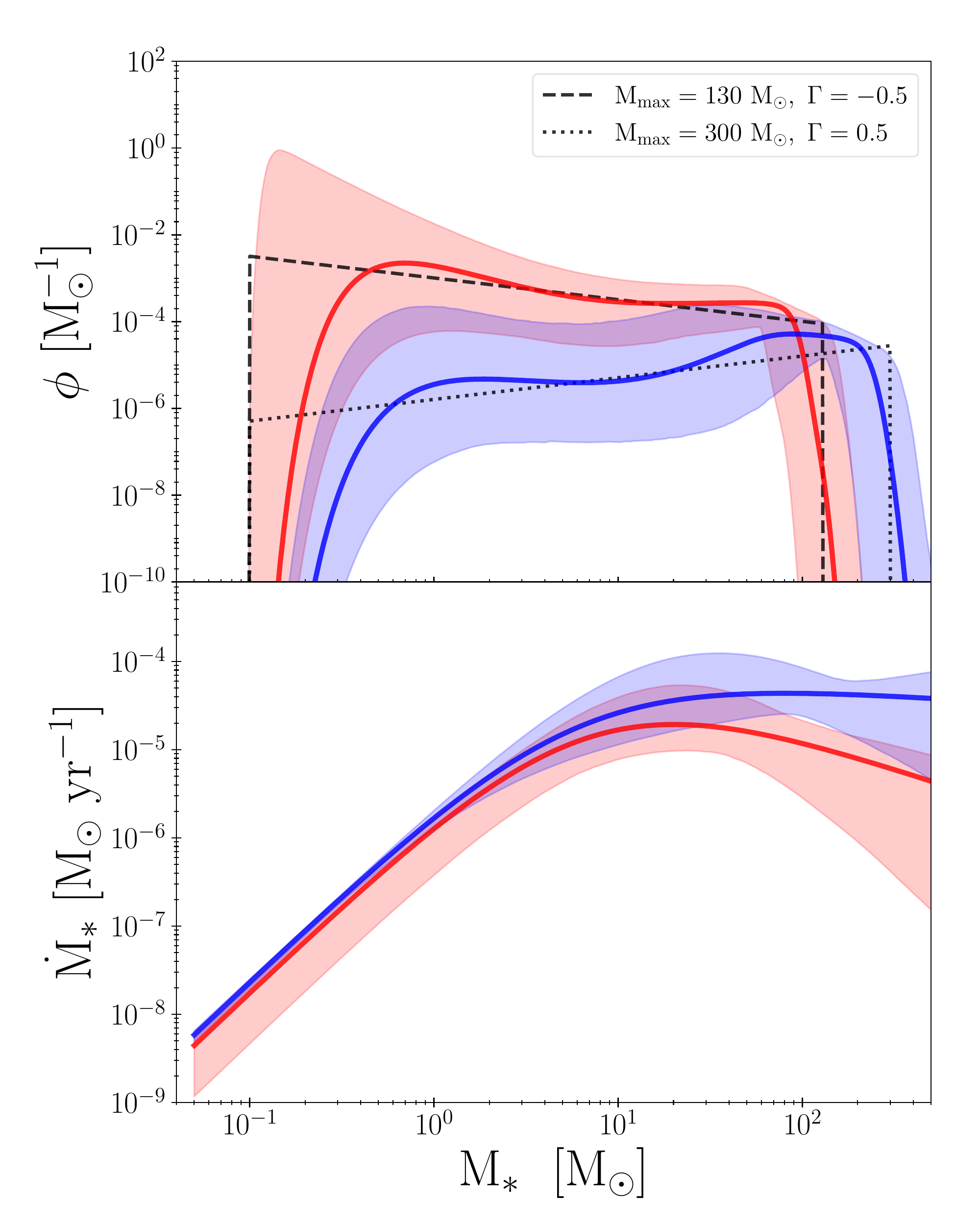}
\end{center}
\caption{
{\it Top panel}: the stellar mass function obtained for the IMF-A (red) and IMF-B (blue) from the MCMC realizations so that the high abundance ratio of $\femg=0.2$ is reproduced.
The solid curves represent the best-fit models with the $3\sigma$ intervals of the generated solutions (shaded regions).
The reference IMF models are overlaid with the dashed and dotted curve.
{\it Bottom panel}: 
the $\ms$--$\mdot$ relation to reproduce the observed high abundance ratio of $\femg\simeq 0.2$, requiring stellar Bondi-like accretion that is regulated by stellar feedback at $\ms >10~\msun$.
}
\label{fig:imf}
\end{figure}

We perform this fitting procedure for the two reference IMF models shown in Table~\ref{table:model}.
The top panel of Figure~\ref{fig:imf} presents the stellar mass distribution obtained for the IMF-A (red) 
and IMF-B (blue) from the MCMC realizations.
For both the cases, the generated mass functions have two peaks at $\ms \lesssim 1$ and $\gtrsim 100~\msun$, 
but the best-fit solutions (solid curves) reproduce shapes consistent with the reference IMFs.
The $\ms$--$\mdot$ relation generated by the fitting method is shown in the bottom panel of Figure~\ref{fig:imf}.
The mean values with standard deviations of $\mdotmin$, $\mc$ and $\alpha$, are also listed in Table~\ref{table:model}.
In the low mass regime ($\ms < 10~\msun$), the mass growth rates for the two cases are almost identical.
Remarkably, the normalization of the mass growth rate at $\ms \lesssim 1~\msun$ appears to be consistent with the Bondi accretion rate onto a star embedded in star-forming regions of an AGN disk, where the typical density and temperature of the gas is 
$n \sim 10^{7 \mathcal{-}9}~{\rm cm^{-3}}$ and gas temperature $T \sim 10^{3 \mathcal{-}4}~{\rm K}$, respectively 
\citep[e.g.,][]{Cantiello2021ApJ}.
Above $\ms \sim 10~\msun$, the best-fit models require the mass accretion rate to decrease with stellar mass.
The critical mass for the transition is consistent with that predicted by analytical models for stellar growth in an AGN disk, 
mainly owing to stellar feedback by radiation and/or winds and 
the tidal effect caused by the central SMBHs \citep[e.g.,][]{Fukushima2018MNRAS, Dittmann2021arXiv}.
Moreover, at this mass range, the IMF-B requires a weaker suppression in $\mdot$ (i.e., a higher value of $\alpha$) to reproduce a larger fraction of stars with $\ms > 100~\msun$.
This implies that the IMF slope and maximum mass depend sensitively on the nature of stellar mass growth and self-regulation mechanisms in the nuclear regions.
Despite a large uncertainty of the power-law index $\alpha$ based on theoretical studies, our MCMC procedure suggests that a reasonable range of $-1.5 \lesssim \alpha \lesssim 0$ successfully reproduces the reference top-heavy IMFs.

In summary, the top-heavy IMF that reproduces the chemical abundance of BLRs can be reasonably 
explained with the star-forming AGN disk model.
We also claim that the $\ms$--$\mdot$ relation obtained in this paper gives a meaningful comparison to 
future theoretical studies for the stellar mass growth mechanism around AGNs.

\section{Black hole merger rates in AGN disks}\label{sec:bhmr}

As shown in Section~\ref{sec:femg}, the high $\femg$ abundance ratio seen in quasars implies that massive stars 
and remnant BHs are born more abundantly than those in normal star-forming environments.
Stellar-mass BHs in AGN disks are expected to migrate inward due to gaseous dynamical friction and form close binaries 
with other BHs, and a fraction of them would merge via GW emission within a Hubble time.
In this section, we quantitatively examine the merger rate of binary BHs (BBHs) that form in the galactic nuclei and 
compare it with the event rate inferred by LIGO/Virgo observations.
In the following discussion, we consider the IMF-A as an example that leads to $\femg > 0.2$ and argue the quantitative difference in the BBH merger rate from the IMF-B case.

The merger rate of BBHs formed in an AGN disk (per unit time) is calculated by
\begin{eqnarray}\label{eq:Rate}
R_{\rm BBH} = \bar{\epsilon}_{\rm m}~ \bar{f}_{\rm bh}~{\rm SFR}_{\rm AGN},
\end{eqnarray}
where $\bar{\epsilon}_{\rm m}$ is the BH merger fraction, $\bar{f}_{\rm bh}$ is the number fraction of stars that end up as massive BHs for a given stellar mass function, and ${\rm SFR}_{\rm AGN}$ is the star formation rate (SFR) in the AGN disk.
We set $\bar{\epsilon}_{\rm m} = 0.5$ based on a recent study modeling binary hardening processes in AGN disks
\citep[][]{Tagawa2020ApJ} and assume that those binaries coalesce immediately in a delay time much shorter than a Hubble time.
To evaluate the SFR in the AGN, we suppose that the surface density of SFR is approximated as a single power-law distribution of $\dot{\Sigma}_{\rm \ast} = r^{-\gamma}$ and thus the SFR within a radius of $r$ is given by ${\rm SFR}(<r) \propto r^{2-\gamma}$ for $\gamma \neq 2$.
Observationally, the SFR within $\sim 100~\pc$ is known to correlate with the BH accretion rate $\dot{M}_{\rm AGN}$ estimated 
from the radiative luminosity by assuming a 10\% radiative efficiency ($\eta_{\rm rad}=0.1$) and the ratio is 
$\dot{M}_{\rm AGN}/{\rm SFR(<100~\pc)} \sim O(0.1)$ \citep[][]{Esquej2014ApJ}. 
Using the empirical relation, we estimate the SFR in the AGN disk as 
${\rm SFR}_{\rm AGN} \sim 10\dot{M}_{\rm AGN} (r_{\rm d}/100~\pc)^{2-\gamma}$, where $r_{\rm d}$ is the size of the AGN disk.
We set the AGN disk size to $r_{\rm d} = 10~\pc$, within which BHs can migrate inward and merge with other ones within $10^8$ yr, the typical lifetime of AGN disks \citep[e.g,][]{Stone2017MNRAS, Dittmann2020MNRAS}.
In this calculation, we assume $\gamma = 1$, leading to ${\rm SFR}_{\rm AGN} \sim \dot{M}_{\rm AGN} (r_{\rm d}/10~\pc)$,
and thus consider that the cosmic star formation rate within $r_{\rm d}$ equals to the cosmic mass accretion rates onto the SMBH measured by \citet{Delvecchio2014MNRAS}.
This is also motivated by the fact that star-forming galaxies tend to have nuclear stellar clusters 
as massive as the central SMBHs within a few pc from the galactic centers \citep[][]{Georgiev2016MNRAS}.
Note here that with a larger value of $\gamma$, the SFR is weighted in the inner part of the AGN disk.
We discuss how the choice of the index $\gamma$ affects the estimate of BH merger rates below.

Next, we divide the AGN disk into two regions.
In the inner region at $r < r_{\rm in} = 0.1~\pc$, which corresponds to the typical size of the BLRs of AGNs 
\citep[e.g,][]{Peterson1993PASP, Kaspi2000ApJ}, 
we assume that the stellar mass distribution follows the IMF-A owing to efficient stellar growth in the AGN disk 
\citep{Thompson2005ApJ, Cantiello2021ApJ}.
As argued in Section~\ref{sec:femg}, the mass function shape is adopted so that the $\femg$ abundance ratio 
becomes as high as $\simeq 0.2$.
In the outer region at $r_{\rm in} \leq  r \leq r_{\rm d}$, we consider that the stellar population follows a Salpeter IMF 
characterized by $\Gamma = -2.35$ and $\mmax = 50~\msun$.
Under these conditions, we calculate the BH formation rate in the AGN disk as 
\begin{align}
\bar{f}_{\rm bh}~{\rm SFR}_{\rm AGN} &= f_{\rm bh, in} {\rm SFR}(<r_{\rm in}) 
+ f_{\rm bh, out} [{\rm SFR}_{\rm AGN} - {\rm SFR}(<r_{\rm in})],\nonumber\\
& \simeq \left[\frac{f_{\rm bh, in}}{f_{\rm bh, out}}\left(\frac{r_{\rm in}}{r_{\rm d}}\right) + 1 \right] f_{\rm bh, out} \dot{M}_{\rm AGN}.
\label{eq:BHMR}
\end{align}
The number fraction of BHs is calculated by integrating the IMF over a mass range where BHs are left behind.
The stellar yield model we adopt in this paper predicts that only 10\% of the initial 
progenitor mass end up in their compact remnants as a consequence of enormous iron 
ejection from their collapsing cores \citepalias{Umeda2008ApJ, Nomoto2013ARA&A}.
We here assume that the minimum stellar mass leaving a BH is $\ms = 30~\msun$ and thus 
the minimum BH mass is $M_{\rm \bullet, min} = 3~\msun$ \citep[e.g.,][]{Ozel2010ApJ, Spera2015MNRAS}.
Massive stars with $\ms = 140$--$260~\msun$ are considered to take place PISNe without forming compact remnants.
Note that while stars with $\ms > 260~\msun$ would leave BHs, the IMF-A and the Salpeter IMF we consider do not cover such an extremely massive population.
Thus, over the mass range of $30\leq \ms/\msun \leq 140$, we obtain $f_{\rm bh, in} = 1.2 \times 10^{-2}~\msun^{-1}$ and $f_{\rm bh, out} = 6.6 \times 10^{-4}~\msun^{-1}$ in the inner and outer region of the AGN disk, respectively.
It is also worth noting that the 10\% of the converting fraction associated with the HN and SLSN models leads to the mean remnant BH mass of $\bar{M_\bullet} = 7.4~\msun$ $(3.8~\msun)$ in the inner (outer) region.
These values are compatible with the low-mass peak in the mass spectra of merging BHs inferred from LIGO-Virgo observations, but do not explain the rate for heavier BBHs at $\mbh \sim 30~\msun$ \citep{LV2021}.
Therefore, further mass growth processes of remnant BHs, such as continuous gas accretion and repeated mergers in dense AGN disks, are essential to simultaneously reproduce the chemical abundance ratio in BLRs and the mass distribution of merging BHs \citep[][]{Inayoshi2016MNRAS, Safarzadeh2020ApJ, Tagawa2020ApJ, Tagawa2021arXiv, Wang2021ApJ}.

Finally, combining Eqs.~(\ref{eq:Rate}) and (\ref{eq:BHMR}), the cosmic merger rate of BBHs (in units of ${\rm Gpc^{-3}~yr^{-1}}$) is calculated as
\begin{eqnarray}
\mathcal{R}_{\rm BBH} = \bar{\epsilon}_{\rm m}
\left[\frac{f_{\rm bh, in}}{f_{\rm bh, out}}\left(\frac{r_{\rm in}}{r_{\rm d}}\right) +1 \right] f_{\rm bh, out} \Psi_{\rm AGN},
\label{eq:R_bbh}
\end{eqnarray}
where the mass accretion rate onto an SMBH in an AGN disk is replaced with the cosmic BH accretion rate density (BHAD) estimated from the AGN bolometric luminosity function of $dN/dL_{\rm AGN}$ as
\begin{eqnarray}
\Psi_{\rm BHAD}(z) = \int_0^\infty \frac{1-\eta_{\rm rad}}{\eta_{\rm rad}c^2}\frac{dN(z,L_{\rm AGN})}{d\ln L_{\rm AGN}}dL_{\rm AGN}.
\end{eqnarray}
We here adopt the function form of $\Psi_{\rm BHAD}(z)$ quantified by \citet{Delvecchio2014MNRAS}, where the AGN bolometric luminosity is calculated by 
a broad-band spectral energy distribution decomposition for obscured AGNs at $0<z<4$ found by Herschel and the Wide-field Infrared Survey Explorer (WISE).

\begin{figure}
\begin{center}
\includegraphics[width=\columnwidth]{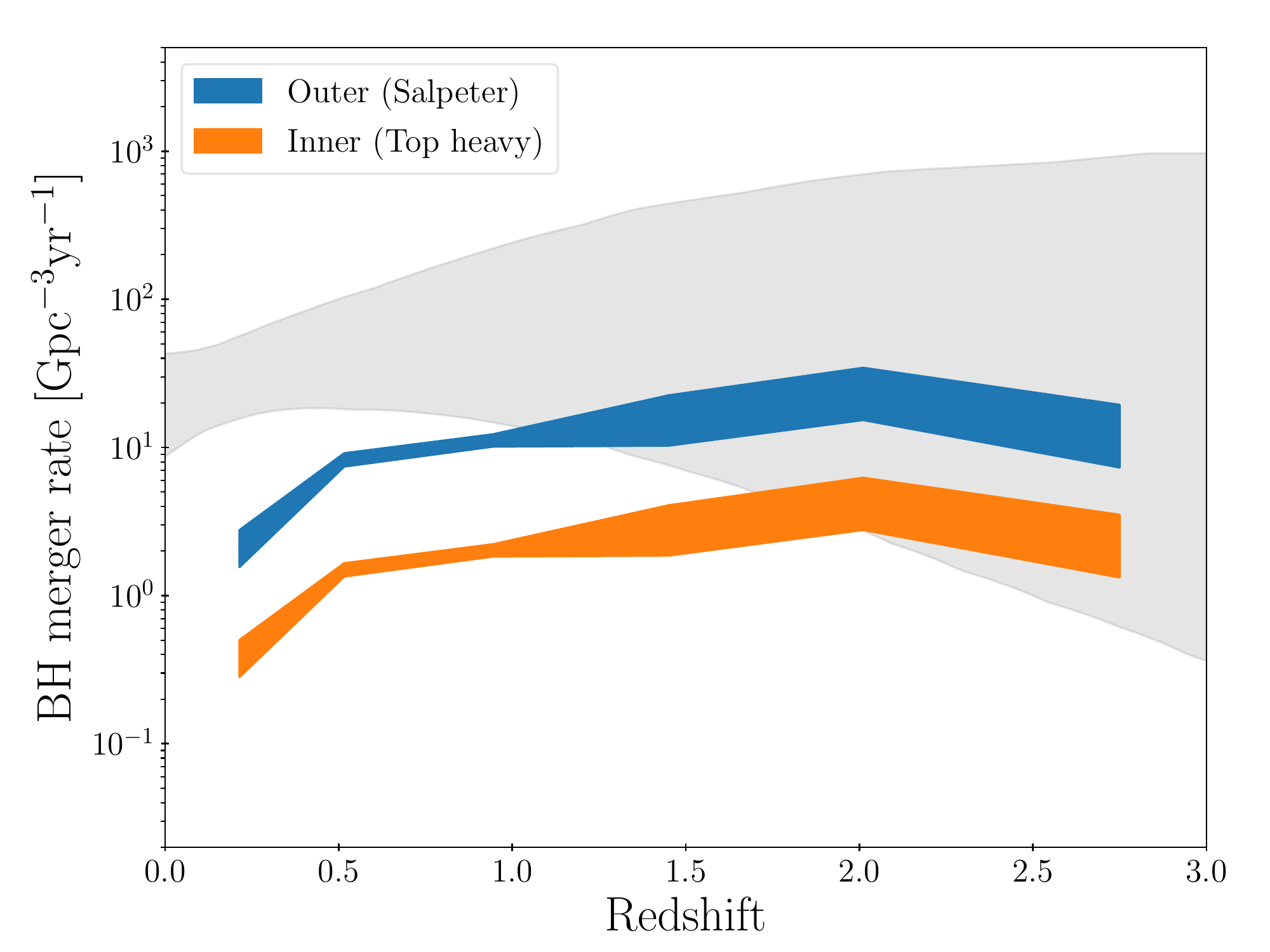}
\end{center}
\caption{
The cosmic merger rates of massive BBHs formed in AGN disks.
The blue and orange curves represent the merger rates originating from the inner ($r < 0.1~\pc$) and the outer ($0.1~\pc < r < 10~\pc$) region of AGN disks, for which the top-heavy and the Salpeter IMF are assumed, respectively.
The variation of the rate at the same redshift is caused by the uncertainty in the cosmic mass accretion rates onto the SMBHs measured by \citet{Delvecchio2014MNRAS}. 
The BBH merger rate inferred from the LIGO-Virgo observations is shown with the grey shaded area \citep{LVK2021a}.
}
\label{fig:rate}
\end{figure}

Figure~\ref{fig:rate} shows the redshift-dependent BH merger rate at $0<z<3$ originating from the inner (orange) and outer (blue) regions of AGN disks.
Since the two rates scale with the cosmic BHAD, the merger rates peak at the cosmic noon at $z\sim 2$ and decline toward lower redshifts.
In the high-redshift universe ($z > 1$), the BBH merger rate associated with the AGN outer regions is within the range inferred from GW observations in the LIGO/Virgo O3a run \citep[the gray region;][]{LVK2021a}.
On the other hand, the BBH merger rate in the AGN inner regions, where the BBH formation efficiency is higher owing to a top-heavy mass distribution but the total SFR in the smaller area is lower, is about five times lower than that in the outer regions.
This overall trend holds as long as the index is $\gamma \lesssim 1.5$.
At $z<0.5$, the contribution from the two populations would explain about 10~\% of the local merger rate of $\mathcal{R}_{\rm BBH}\simeq 19.1~{\rm Gpc}^{-3}~{\rm yr}^{-1}$ \citep{LVK2021a}.
Although the merger rates we calculate are lower than the LIGO/Virgo prediction in the local universe, this discrepancy could be resolved by taking into account a further delay-time of BH mergers, i.e., allowing a fraction of those BHs to coalesce even after host AGN disks have diminished.

Here, we briefly argue the case with the IMF-B.
As shown in Table~\ref{table:model}, 
the number fraction of BHs owing to stars with $\ms = 30$--$140~\msun$ for the IMF-B is 4 times lower than that for the IMF-A 
because the total number of stars per unit mass is lower for a top-heavier mass distribution.
Therefore, the BBH merger rate in the inner part of AGN disks scales down accordingly from the rate expected for the IMF-A case in Figure~\ref{fig:rate}.
In the case of the IMF-B, we also consider BH formation from massive stars with $\ms > 260~\msun$, where ejection of stellar yields is negligible in the model we adopt. 
Therefore, the mass of BH remnants becomes as high as the progenitor mass in the pre-collapsing phase.
As a result, intermediate massive BHs (IMBHs) with $\mbh \gtrsim 100~\msun$ is considered to form from massive stars with $\ms > 260~\msun$, and the number fraction of IMBHs is estimated as $f_{\rm imbh} = 1.1 \times 10^{-3}~\msun^{-1}$.
This value leads to the cosmic merger rate of $\mathcal{R}_{\rm BBH} \sim 0.04~{\rm Gpc}^{-3}~{\rm yr}^{-1}$,
which is slightly lower than the upper limit of IMBH merger rates of $0.056~{\rm Gpc}^{-3}~{\rm yr}^{-1}$ constrained by the LIGO/Virgo observations \citep[][]{LVK2021b}.

The model with the IMF-B yields another notable difference from the IMF-A case
in the event rate of PISNe, whose progenitors are considered to be 
as massive as $\ms = 140-260~\msun$.
Therefore, a stellar population that follows the IMF-B leads to a larger number of PISNe at a fraction of 
$f_{\rm pisn} = 2.7 \times 10^{-3}~\msun^{-1}$ (the rightmost column in Table~\ref{table:model}), while 
the IMF-A with $M_{\rm max}=130~\msun$ does not produce a PISN ($f_{\rm pisn}=0$).
Replacing $f_{\rm bh}$ with $f_{\rm pisn}$ in Eq. (\ref{eq:R_bbh}) and setting $\bar{\epsilon}_{\rm m} = 1$, we estimate 
the PISN rate in AGN disks as $R_{\rm pisn} \sim 0.1~{\rm Gpc^{-3}~yr^{-1}}$ at $z = 0.2$ and $\sim 3~{\rm Gpc^{-3}~yr^{-1}}$ at $z = 2$.
Those expected rates are consistent with the fact that there has not been a clear detection of PISNe.
This gives an upper limit of the event rate at $\sim 1~{\rm Gpc^{-3}~yr^{-1}}$ in the local universe \citep[][]{Pan2012MNRAS}.
It is worth noting that the upcoming observations with Vela C. Rubin Observatory's LSST will enable us to explore PISN events up to $z \sim 2$ \citep[][]{Kasen2011ApJ}, and could detect a few hundreds of PISNe per year if massive stars universally formed in AGNs following extremely top-heavy IMFs.
Therefore, future observations of luminous transient events would be useful to constrain the stellar IMF in AGN disks.

\section{Other explanations of high iron to magnesium abundance ratio}\label{sec:other}

In this section, we briefly mention two other possible scenarios that a high abundance ratio of $\femg$ is achieved in AGN disks.
The first one to increase the iron abundance of the interstellar medium is associated with type Ia SNe produced by thermonuclear explosions of white dwarfs (WDs) that exceed the Chandrasekhar mass limit owing to mass accretion through the AGN disk.
If the growth timescale of WDs is fast enough to produce type Ia SNe in a shorter delay time since star formation ($\ll 1$ Gyr),  
the nuclear regions of AGNs would be quickly enriched by the yields composed of iron, enabling us to explain the high abundance 
ratio of $\femg$ in quasars observed at $z>6$.
For instance, the $\ms$--$\mdot$ relation shown in Fig. \ref{eq:mdot} yields $\mdot \sim 10^{-6}~\msun~\yr^{-1}$ at $\ms = 1~\msun$, implying that WDs in a star forming AGN disk grow to the Chandrasekhar mass within a few Myr.
However, the progenitors of those WDs would grow further in mass and become heavy enough to end up as NSs or BHs instead of WDs \citep[e.g.,][]{Cantiello2021ApJ, Dittmann2021arXiv}.
Moreover, an analytical study by \citet{Pan2021arXiv} claimed that accreting WDs spin up and thus prevent mass accretion before reaching the critical mass for the onset of thermonuclear explosions.
Therefore, type Ia SNe induced by accreting WDs in dense AGN disks would not make a significant contribution to iron-enrichment of BLRs.

Another possible production of iron is through GRB events caused by massive stellar progenitors of $\ms > 260~\msun$.
While such massive stars are considered to directly collapse to BHs at the end of their lifetime without producing iron-rich yields,
relativistic jets associated with GRB-like explosions, if any, would release a large fraction of newly synthesized metals into the interstellar medium.
\citet{Ohkubo2006ApJ} has conducted nucleosynthesis calculations for  massive stars with $\ms = 500$ and $1000~\msun$ and found that massive GRB-like explosions preferentially produce iron-rich ejecta with $\femg \sim 0.5$.
Therefore, an extremely top-heavy IMF that leads to a large number of massive GRBs would be a possible solution to reproduce the high abundance ratio of $\femg > 0.2$ seen in high-$z$ quasars.
However, the stellar yield model owing to massive GRBs depends sensitively on the explosion energy and the jet opening angles.
We leave this issue for future investigation and make a robust conclusion on chemical enrichment in quasars using yield models calibrated by various SN observations.


\section{Summary}\label{sec:summary}

In this paper, we argue the properties of the stellar mass distribution in galactic nuclear regions to explain 
the high iron to magnesium abundance ratio of $\femg \gtrsim 0.2$, which is observed in broad-line regions (BLRs) 
of active galactic nuclei (AGNs).
Nuclear synthesis models suggest that massive explosive events such as super-luminous supernovae (SLSNe) 
with progenitor mass of $\ms \sim 60$--$140~\msun$ and pair-instability supernovae (PISNe) with $\ms \sim 220$--$260~\msun$ 
produce highly iron-enriched ejecta and effectively enhance $\femg$ of the interstellar medium.
We calculate the mass-integrated stellar yields varying the maximum mass $\mmax$ and power-law index $\Gamma$ 
of the stellar initial mass function (IMF), and find that the iron-enriched environments with $\femg \gtrsim 0.2$ form when 
the IMF is characterized with $\mmax \sim 100$--$150~\msun$ and $\Gamma \gtrsim -1$ or $\mmax \gtrsim 250~\msun$ and $\Gamma \gtrsim 0$.
This suggests that massive stars preferentially form in galactic nuclei and promote chemical enrichment in BLRs.

Next, we discuss how those top-heavy IMFs are established in the galactic nuclear regions.
Following a theoretical model of massive star formation via rapid accretion through dense AGN disks, 
we calculate the time-evolution of the stellar mass distribution and model the stellar mass growth rate so that 
the abundance ratio of $\femg \gtrsim 0.2$ observed in AGNs is explained by their explosive events
(see Eqs. \ref{eq:cont} and \ref{eq:mdot}).
The model requires Bondi-like stellar growth that is regulated by radiative/mechanical stellar feedback at $\ms > 10~\msun$.

Finally, we calculate the cosmic merger rate of binary BHs formed in AGN disks, assuming the top-heavy IMF in the inner BLRs ($r < 0.1~\pc$) to reproduce $\femg > 0.2$ and a Salpeter IMF in the outer part ($0.1~\pc < r < 10~\pc$).
We find that the merging BH population in the outer region is responsible for the total merger rate, which is generally consistent with the rate inferred by the LIGO and Virgo observations at $z \sim 1$--$3$.
The BH merger rate in AGN disks declines toward lower redshifts, as the cosmic AGN activity fades out, but still accounts for about 10~\% of the total merger rate observed in the local universe.
We also suggest that top-heavy IMFs achieved in the inner region do not necessarily promote formation of massive BHs with $\mbh \gtrsim 10~\msun$ since enormous mass ejection from their HNe and SLSNe is required to reproduce the high iron to magnesium abundance ratio in BLRs.
Thus, further mass growth mechanisms of remnant BHs, such as continuous gas accretion and repeated BH mergers in AGN disks, would be essential to establish a high-mass peak around $\mbh \sim 30~\msun$ in the observed mass spectra of merging BHs.
In addition, the top-heavy hypothesis would require a high detection rate of PISNe at $z \lesssim 2$ in the upcoming LSST observations.
This will enable us to put a meaningful constraint on the stellar mass distribution in the nuclear regions of AGNs.


\section*{Acknowledgements}

We thank K.~Hotokezaka and M.~Onoue for useful discussions. 
This work was supported in part by JSPS KAKENHI Grant Numbers 17H06363, 17K14249, 20H00179, 20H05855, 21H04499, 21H04997, and 21K20378.
K. I. acknowledges support from the National Natural Science Foundation of China (12073003, 12003003, 11721303, 11991052, 11950410493), the National Key R\&D Program of China (2016YFA0400702).


\section*{Data availability}

The data underlying this article will be shared on reasonable request to the corresponding author.


\appendix

\section{Yield table}\label{sec:table}

\begin{table}
\begin{center}
\caption{The yield table adopted in this paper. At $\ms = 60$--$140~\msun$, the naked (bracketed) values represent the yields with (without) considering SLSN-like explosive events in \citetalias{Umeda2008ApJ} (\citetalias{Nomoto2013ARA&A}).}
\label{table:yield}
\begin{tabular}{cccc} \hline \hline
$\ms~[\msun]$ & $y_{\rm Mg}~[\msun]$ & $y_{\rm Fe}~[\msun]$ & $\femg$ \\
\hline
11 & 0.028 & 0.01 & $-0.23$ \\
13 & 0.086 & 0.07 & $-0.047$ \\
15 & 0.068 & 0.07 & $0.0058$ \\
18 & 0.16 & 0.07 & $-0.18$ \\
20 & 0.17 & 0.082 & $-0.16$ \\
25 & 0.15 & 0.096 & $-0.1$ \\
30 & 0.22 & 0.16 & $-0.069$ \\
40 & 0.34 & 0.26 & $-0.062$ \\
50 & 0.5 & 0.34 & $-0.086$ \\
60 & 0.95 (0.66) & 3.2 (0.42) & $0.27$ ($-0.099$) \\
80 & 1.3 (0.99) & 5 (0.59) & $0.3$ ($-0.11$) \\
100 & 1.3 (1.3) & 6.4 (0.76) & $0.35$ ($-0.12$) \\
140 & 1.3 (1.7) & 6.4 (0.95) & $0.35$ ($-0.13$) \\
141 & 1.5 & 1.3e-13 & $-6.7$ \\
150 & 3 & 0.012 & $-1.2$ \\
159 & 3.5 & 0.1 & $-0.79$ \\
167 & 3.7 & 0.13 & $-0.75$ \\
176 & 4 & 0.41 & $-0.51$ \\
185 & 4.2 & 1.3 & $-0.26$ \\
193 & 4.4 & 3 & $-0.086$ \\
202 & 4.4 & 5.8 & $0.062$ \\
211 & 4.4 & 9.6 & $0.17$ \\
220 & 4.3 & 14 & $0.27$ \\
228 & 4.5 & 19 & $0.32$ \\
237 & 4.5 & 25 & $0.38$ \\
246 & 4.4 & 32 & $0.44$ \\
260 & 4.4 & 40 & $0.49$ \\
 \hline \hline \\
\end{tabular}
\end{center}
\end{table}



\bibliographystyle{mnras}
\bibliography{refs.bib} 



\bsp	
\label{lastpage}
\end{document}